\documentclass{article}
\usepackage{frascatiphys_SP}
\newcommand{\mc}{\multicolumn}
\newcommand{\gsim}{\mathrel{\mathop{\kern 0pt \rlap
  {\raise.2ex\hbox{$>$}}}
  \lower.9ex\hbox{\kern-.190em $\sim$}}}
\begin{document}
\title{ 
TOWARDS A SOLUTION OF THE $\Gamma_{\rm n}/\Gamma_{\rm p}$ PUZZLE
IN THE WEAK DECAY OF $\Lambda$--HYPERNUCLEI
}
\author{
W.~M.~Alberico$^1$, \underline{G.~Garbarino}$^1$, A. Parre\~no$^2$ and A. Ramos$^2$ \\
 \\
\em $^1$Dipartimento di Fisica Teorica, Universit\`a di Torino
and INFN, \\ Sezione di Torino, I--10125 Torino, Italy \\
$^2$Departament d'Estructura i Constituents de la Mat\`{e}ria, \\
Universitat de Barcelona, E--08028 Barcelona, Spain}

\maketitle
\baselineskip=11.6pt
\vskip 23mm
%
\baselineskip=14pt
%
%
\noindent {\bf Introduction} ---
For many years, a theoretical explanation of the large
experimental values of the ratio, $\Gamma_n/\Gamma_p$,
between the neutron-- and proton--induced non--mesonic
decay widths, $\Gamma(\Lambda n\to nn)$ and $\Gamma(\Lambda p\to np)$,
of $\Lambda$--hypernuclei has been missing\cite{Al02}.

In this contribution we discuss some results of a calculation\cite{ours} of 
nucleon--nucleon coincidence distributions for the hypernuclear non--mesonic weak decay (NMWD).
The work is motivated by the fact that correlation observables are expected
to allow a cleaner extraction of $\Gamma_n/\Gamma_p$ from data
than single--nucleon observables. Moreover, coincidence experiments
have been performed recently at KEK\cite{outa}.

A one--meson--exchange model for the $\Lambda N\to nN$ processes in finite nuclei 
has been combined with an intranuclear
cascade code, which takes into account the nucleon final state interactions (FSI).
The $\Lambda NN\to nNN$ process is included by treating the nuclear finite
size effects via a local density approximation scheme. 
For details on the models employed see Ref.\cite{ours}.

Preliminary results for the angular asymmetries in the
NMWD of polarized $\Lambda$--hypernuclei are also presented.

\vskip 2mm
\noindent {\bf Coincidence observables and determination of $\Gamma_n/\Gamma_p$} ---
The ratio, $N^{\rm wd}_{nn}/N^{\rm wd}_{np}$, between the number of weak decay 
$nn$ and $np$ pairs equals $\Gamma_n/\Gamma_p$. 
Due to FSI and two--body induced decays, one predicts:
\begin{equation}
\label{ratio-nn}
\frac{\Gamma_n}{\Gamma_p}\equiv \frac{N^{\rm wd}_{nn}}{N^{\rm wd}_{np}}
\neq \frac{N_{nn}}{N_{np}}\equiv 
R_2\left[\Delta \theta_{12}, \Delta T_n, \Delta T_p\right] ,
\end{equation}
when the observable numbers
$N_{nn}$ and $N_{np}$ are determined by employing particular
pair opening angle and nucleon kinetic energy intervals.
The results of Ref.\cite{ours} clearly show the dependence of $N_{nn}/N_{np}$
on $\Delta \theta_{12}$, $\Delta T_n$ and $\Delta T_p$;	
$N_{nn}/N_{np}$ turns out to be much less sensitive   
to FSI effects and variations of energy
cuts and angular restrictions than $N_{nn}$ and $N_{np}$ separately.

The numbers of nucleon pairs $N_{NN}$ ---which we consider to be
normalized per NMWD--- are related to the corresponding quantities for the neutron 
($N^{\rm 1Bn}_{NN}$) proton ($N^{\rm 1Bp}_{NN}$) and 
two--nucleon ($N^{\rm 2B}_{NN}$) induced processes by:
\begin{equation}
\label{1-2}
N_{NN}=\frac{N^{\rm 1Bn}_{NN}\, \Gamma_n+ N^{\rm 1Bp}_{NN}\, \Gamma_p+N^{\rm 2B}_{NN}\, \Gamma_2}
{\Gamma_n +\Gamma_p+\Gamma_2}
\equiv N^{\Lambda n\to nn}_{NN}+ N^{\Lambda p\to np}_{NN}+
N^{\Lambda np\to nnp}_{NN} , \nonumber
\end{equation}
where $N^{\rm 1Bn}_{NN}\equiv N^{\Lambda n\to nn}_{NN} 
(\Gamma_n +\Gamma_p+\Gamma_2)/\Gamma_n$, etc.

In Table \ref{ome-che} the ratios $N_{nn}/N_{np}$ 
predicted by the one--pion--exchange (OPE) and one--meson--exchange models 
(OMEa and OMEf, using NSC97a and NSC97f potentials, respectively) for $^5_\Lambda$He
and $^{12}_\Lambda$C are given for the back--to--back kinematics
($\cos \theta_{NN}\leq -0.8$) and nucleon kinetic energies 
$T_n, T_p\geq 30$ MeV. The OMEa and OMEf results
are in agreement with the preliminary KEK data: this comparison
provides an indication for 
a ratio $\Gamma_n/\Gamma_p\simeq 0.3$ in both hypernuclei.
\begin{table}
\begin{center}
\caption{Predictions for $R_2\equiv N_{nn}/N_{np}$
in $^5_\Lambda$He and $^{12}_\Lambda$C.
The (preliminary) data are from KEK--E462 
and KEK--E508\protect\cite{outa}.}
\label{ome-che}
\begin{tabular}{c|c c | c c}
\hline
\mc {1}{c|}{} &
\mc {1}{c}{$^5_\Lambda$He} &
\mc {1}{c|}{} &
\mc {1}{c}{$^{12}_\Lambda$C} &
\mc {1}{c}{} \\ 
Model   & $N_{nn}/N_{np}$ & $\Gamma_n/\Gamma_p$ & $N_{nn}/N_{np}$ & $\Gamma_n/\Gamma_p$ \\ \hline
OPE     & $0.25$  & $0.09$  & $0.24$ & $0.08$  \\
OMEa    & $0.51$  & $0.34$  & $0.39$ & $0.29$  \\
OMEf    & $0.61$  & $0.46$  & $0.43$ & $0.34$ \\ \hline
EXP  & $0.44\pm 0.11$  &  &   $0.40\pm 0.09$       &  \\ \hline
\end{tabular}
\end{center}
\end{table}

We have then performed a weak--decay--model independent analysis
of KEK coincidence data. 
The 6 weak--decay--model independent quantities
$N^{\rm 1Bn}_{nn}$, $N^{\rm 1Bp}_{nn}$, $N^{\rm 2B}_{nn}$,
$N^{\rm 1Bn}_{np}$, $N^{\rm 1Bp}_{np}$ and $N^{\rm 2B}_{np}$
of Eq.~(\ref{1-2}) are used to evaluate $\Gamma_n/\Gamma_p$ as:   
\begin{equation}
\label{fit}
\frac{\Gamma_n}{\Gamma_p}=
\frac{\displaystyle N^{\rm 1Bp}_{nn}+N^{\rm 2B}_{nn} \frac{\Gamma_2}{\Gamma_1}
-\left(N^{\rm 1Bp}_{np}+N^{\rm 2B}_{np} \frac{\Gamma_2}{\Gamma_1}
\right)\frac{N_{nn}}{N_{np}}}
{\displaystyle \left(N^{\rm 1Bn}_{np}+N^{\rm 2B}_{np}
\frac{\Gamma_2}{\Gamma_1} \right) \frac{N_{nn}}{N_{np}}
-N^{\rm 1Bn}_{nn}-N^{\rm 2B}_{nn}
\frac{\Gamma_2}{\Gamma_1}} ,
\end{equation}
from appropriate $\Gamma_2/\Gamma_1$ values.
By using the KEK data of Table~\ref{ome-che} we obtain:
\begin{equation}
\label{fit1}
\frac{\Gamma_n}{\Gamma_p}\left(^5_\Lambda {\rm He}\right)=0.39\pm0.11\,\,\,
{\rm if}\,\,\, \Gamma_2=0\,\, , 
\frac{\Gamma_n}{\Gamma_p}\left(^5_\Lambda {\rm He}\right)=0.26\pm0.11\,\,\,
{\rm if}\,\,\, \frac{\Gamma_2}{\Gamma_1}=0.2 ,
\end{equation}
\begin{equation}
\frac{\Gamma_n}{\Gamma_p}\left(^{12}_\Lambda {\rm C}\right)=0.38\pm 0.14\,\,\,  
{\rm if}\,\,\, \Gamma_2=0\,\, ,     
\frac{\Gamma_n}{\Gamma_p}\left(^{12}_\Lambda {\rm C}\right)=0.29\pm 0.14\,\,\,    
{\rm if}\,\,\, \frac{\Gamma_2}{\Gamma_1}=0.25 .
\end{equation}
These values are substantially smaller than 
those obtained from single--nucleon spectra analyses and
are in agreement with pure theoretical
predictions\cite{good}. In our opinion, this represents an important 
progress towards the solution of the $\Gamma_n/\Gamma_p$ puzzle.

Forthcoming data from KEK and FINUDA\cite{finuda}
could be directly compared with the results reported here and in Ref.\cite{ours}.
This will permit to achieve better determinations of $\Gamma_n/\Gamma_p$ 
and to establish the first constraints on $\Gamma_2/\Gamma_1$.

\vskip 2mm
\noindent {\bf The asymmetry puzzle} ---
An intriguing open problem concerns
an angular asymmetry in the emission of NMWD protons 
from polarized hypernuclei.
While theory predicts a negative intrinsic $\Lambda$ 
asymmetry $a_\Lambda$, with a moderate dependence on the hypernucleus,
the measurements seem to favor
$a^{\rm M}_\Lambda(^5_\Lambda{\vec {\rm H}{\rm e}})>0$ and
$a^{\rm M}_\Lambda(^{12}_\Lambda{\vec {\rm C}})<0$. 
However, while one predicts
$a_\Lambda(^5_\Lambda{\vec {\rm H}{\rm e}})\simeq a_\Lambda(^{12}_\Lambda{\vec {\rm C}})$,
there is no known reason to expect this approximate equality to be valid
for the observable asymmetry, $a^{\rm M}_\Lambda$. 
To overcome this problem, we are evaluating\cite{asym} the effects of the
nucleon FSI on the NMWD of polarized hypernuclei
and performing the first calculation of $a^{\rm M}_\Lambda$.

In table~\ref{results} we show preliminary OMEf results
for the weak decay and observable proton intensities, 
$I(\theta)=I_0(1+p_\Lambda\, a_\Lambda \cos \theta)$ 
and $I^{\rm M}(\theta)=I^{\rm M}_0(1+p_\Lambda\, a^{\rm M}_\Lambda \cos \theta)$,
respectively, for
$^5_\Lambda {\vec {\rm H}}{\rm e}$ and $^{12}_\Lambda {\vec {\rm C}}$.
As a result of the nucleon FSI, $|a_\Lambda|\gsim |a^{\rm M}_\Lambda|$ for any
value of the proton kinetic energy threshold: when $T^{\rm th}_p=0$,
$a_\Lambda/a^{\rm M}_\Lambda\simeq 2$ for $^5_\Lambda {\vec {\rm H}}{\rm e}$
and $a_\Lambda/a^{\rm M}_\Lambda\simeq 4$ for
$^{12}_\Lambda {\vec {\rm C}}$; $|a^{\rm M}_\Lambda|$ increases
with $T^{\rm th}_p$ and $a_\Lambda/a^{\rm M}_\Lambda\simeq 1$
for $T^{\rm th}_p=70$ MeV in both cases.
\begin{table}
\begin{center}
\caption{Results for the proton intensities
from the NMWD of $^5_\Lambda{\vec {\rm H}{\rm e}}$
and $^{12}_\Lambda{\vec {\rm C}}$.}
\label{results}
\begin{tabular}{l|c c| c c} \hline
\mc {1}{c|}{Model} &
\mc {1}{c}{$^5_\Lambda{\vec {\rm H}}{\rm e}$} &
\mc {1}{c|}{} &
\mc {1}{c}{$^{12}_\Lambda{\vec {\rm C}}$} &
\mc {1}{c}{} \\
                  & $I^{\rm M}_0$    & $a^{\rm M}_\Lambda$   &
$I^{\rm M}_0$    &     $a^{\rm M}_\Lambda$ \\ \hline
{\small Without FSI}                    & $0.69$  & $-0.68$  & $0.75$ & $-0.73$   \\
{\small FSI and $T^{\rm th}_p=0$}       & $1.27$  & $-0.30$  & $2.78$ & $-0.16$  \\
{\small FSI and $T^{\rm th}_p=30$ MeV}  & $0.77$  & $-0.46$  & $1.05$ & $-0.37$  \\
{\small FSI and $T^{\rm th}_p=50$ MeV}  & $0.59$  & $-0.52$  & $0.65$ & $-0.51$  \\
{\small FSI and $T^{\rm th}_p=70$ MeV}  & $0.39$  & $-0.55$  & $0.38$ & $-0.65$  \\ \hline
  KEK\cite{Ma04}  &   & $0.11\pm 0.44$  &  & $-0.44\pm 0.32$  \\ \hline
\end{tabular}
\end{center}
\end{table}
The KEK data quoted in the table correspond to a $T^{\rm th}_p$ varying between     
$30$ and $50$ MeV: our corresponding predictions agree (disagree) with the 
$^{12}_\Lambda{\vec {\rm C}}$ ($^5_\Lambda{\vec {\rm H}}{\rm e}$) datum. 

FSI turn out to be an important ingredient also when studying the NMWD
of polarized hypernuclei, but they cannot 
explain the present asymmetry data. In our opinion, 
new and improved experiments more clearly establishing
the sign and magnitude of $a^{\rm M}_{\Lambda}$ for $s$-- and
$p$--shell hypernuclei are necessary to disclose the origin of the
asymmetry puzzle.




\section*{Acknowledgements}
Work supported by EURIDICE HPRN--CT--2002--00311 and INFN.
Discussions with H. Bhang, T. Maruta, T. Nagae and H. Outa are acknowledged.

\end{document}